\documentclass[a4paper,fleqn,10pt]{article}
\pdfoutput=1
\usepackage{amsmath}
\usepackage{amssymb}
\usepackage{array}
\usepackage{calc}
\usepackage{longtable}
\usepackage{multirow}
\usepackage{slashed}
\usepackage{pstricks}
\usepackage{graphicx}
\usepackage{xspace}
\usepackage{units}
\usepackage{tikz}
\usepackage{textcomp}
\numberwithin{equation}{section}
\usepackage{cite}
\usepackage[pdfborder={0 0 0}]{hyperref}
\usepackage[format=hang,labelfont=bf,hypcap=true]{caption}
\usepackage{subcaption}
\usepackage{sectsty}
\usepackage{enumitem}
\allsectionsfont{\sffamily}
\subsubsectionfont{\mdseries\itshape\large}
\makeatletter
\DeclareRobustCommand*{\bfseries}{%
  \not@math@alphabet\bfseries\mathbf
  \fontseries\bfdefault\selectfont
  \boldmath
}
\makeatother
\let\spreprint\empty
\newcommand{\preprint}[1]{\def\spreprint{\protect#1}}
\let\sinstitute\empty
\newcommand{\institute}[1]{\def\sinstitute{\protect#1}}
\makeatletter
\renewcommand{\maketitle}{\begingroup
  \null\thispagestyle{empty}%
    \ifx\spreprint\empty
      \vskip 5ex
    \else
      \flushright\large\spreprint\vskip 10ex
    \fi
    \vskip 5ex
    \centering
      {\sffamily\bfseries\huge\@title}\vskip 10ex
    \flushleft
      \@author\vskip 2ex
      \ifx\sinstitute\empty
      \else
        {\small\sinstitute}
      \fi
    \vskip 5ex
  \endgroup
}
\makeatother
\renewenvironment{abstract}{\begin{center}
  {\large\sffamily\bfseries Abstract: }
  \begin{minipage}[t]{0.75\textwidth}
}{\end{minipage}\end{center}\vskip 10ex}

\numberwithin{equation}{section}
\allowdisplaybreaks[2]
\newcommand{\MCatNLO}{M\protect\scalebox{0.8}{C}@N\protect\scalebox{0.8}{LO}\xspace}

\newcommand{\LHAPDF}{L\protect\scalebox{0.8}{HAPDF}\xspace}

\newcommand{\Rivet}{R\protect\scalebox{0.8}{IVET}\xspace}

\newcommand{\SmeftDecayLib}{S\protect\scalebox{0.8}{MEFT}D\protect\scalebox{0.8}{ECAY}L\protect\scalebox{0.8}{IB}\xspace}
\newcommand{\Sherpa}{S\protect\scalebox{0.8}{HERPA}\xspace}

\newcommand{\Amegic}{A\protect\scalebox{0.8}{MEGIC}\xspace}
\newcommand{\CSShower}{C\protect\scalebox{0.8}{SSHOWER}\xspace}

\long\def\symbolfootnote[#1]#2{\begingroup%
\def\thefootnote{\fnsymbol{footnote}}\footnote[#1]{#2}\endgroup}

\newcommand{\done}{{\rm d}}

\newcommand{\mc}[1]{\mathcal{#1}}
\newcommand{\mr}[1]{\mathrm{#1}}

\newcommand{\bea}{\begin{eqnarray}}
\newcommand{\eea}{\end{eqnarray}}
\newcommand{\bi}{\begin{itemize}}
\newcommand{\ei}{\end{itemize}}
\newcommand{\hl}{\vphantom{$\int_A^B$}}

\xspace
\xspace
\xspace
\xspace
\xspace
\xspace
\xspace
\xspace
\xspace
\xspace
\xspace
\xspace
\xspace
\xspace
\xspace
\xspace
\xspace
\xspace
\xspace
\xspace
\xspace
\xspace
\xspace
\xspace
\xspace

\newlist{myitemize}{itemize}{3}
\setlist[myitemize]{leftmargin=14em}

\newcolumntype{C}{>{\centering\arraybackslash}p{0.14\textwidth}}

\newlength{\unitcharwidth}
\newcommand{\hc}{\hspace*{\unitcharwidth}}

\hypersetup{
  pdfauthor={Ben Pecjak, Livia E. G. Maskos, Shakeel Ur Rahaman, Marek Schoenherr},
  pdftitle={An event generator for h->bb decays at NLO in SMEFT matched to the parton shower}
}
\preprint{IPPP/26/60 \\MCnet-26-25}
\author{Livia E. G. Maskos, Benjamin D. Pecjak,  Shakeel Ur Rahaman, Marek Sch{\"o}nherr}
\title{An event generator for \texorpdfstring{$h\to b\bar{b}$}{h->bb} decays at NLO in SMEFT matched to a parton shower}
\institute{Institute for Particle Physics Phenomenology, Department of Physics, Durham University, Durham DH1 3LE, United Kingdom\\
$~$\\
Emails:
\href{mailto:ben.pecjak@durham.ac.uk}{\texttt{ben.pecjak@durham.ac.uk}}\,,
\href{mailto:livia.e.maskos@durham.ac.uk}{\texttt{livia.e.maskos@durham.ac.uk}}\,,
\href{mailto:shakeel.u.rahaman@durham.ac.uk}{\texttt{shakeel.u.rahaman@durham.ac.uk}}\,,
\href{mailto:marek.schoenherr@durham.ac.uk}{\texttt{marek.schoenherr@durham.ac.uk}}
}

\def\msbar{$\overline{\hbox{MS}}$}

\begin{document}
\vspace*{10mm}
\maketitle
\vspace*{20mm}
\begin{abstract} 
We present a \textsc{Sherpa} implementation of NLO corrections in dimension-six Standard Model Effective Field 
Theory (SMEFT) to the benchmark process $h\to b\bar b$. The calculation includes NLO QCD corrections matched to
a parton shower using an \MCatNLO-type framework for decays. Virtual weak
effects are included as two-body matrix-element corrections in different
electroweak input schemes, with the corresponding decay events also
evolved with the QCD parton shower. We present numerical results for the inclusive decay rate and 
for differential distributions in the Higgs-boson rest frame, and illustrate how the decay
implementation can be combined with an independently generated production sample by embedding it in 
associated $Zh$ production at the LHC. This proof-of-principle implementation demonstrates how NLO SMEFT
corrections can be incorporated into the widely used \textsc{Sherpa} event generator, providing a basis for extensions to 
other processes and for SMEFT studies with realistic final states and fiducial cuts within that framework.

\end{abstract}
\newpage
\tableofcontents
%= text ===========================================
 
\section{Introduction}
\label{sec:intro}

In light of the current experimental situation at the LHC, where no compelling direct evidence for particles beyond the 
Standard Model has yet emerged, Standard Model Effective Field Theory (SMEFT) has become
a standard framework in which to analyse small deviations from Standard Model (SM) predictions,
see for example~\cite{Brivio:2017vri,Isidori:2023pyp}.
As in the SM itself, phenomenological predictions in SMEFT can be improved by including radiative corrections.   
This has motivated a growing number of complete or partial next-to-leading order (NLO) calculations in dimension-six SMEFT,
mainly for individual processes but also in general-purpose tools.

Given that full NLO SMEFT calculations of even relatively simple processes
are sensitive to a large set of Wilson coefficients, an important application 
is in global fits to a wide variety of measurements of Higgs, top, flavour and electroweak precision observables,
see for example~\cite{Ellis:2020unq,Giani:2023gfq}. 
As such, it is crucial to make  them available in forms that can be used directly in
phenomenological and experimental analyses, including differential
distributions with realistic final states and fiducial selections.  Several
tools and implementations already address various aspects of this
problem.  For instance, \textsc{SMEFT@NLO} automates NLO SMEFT
QCD computations in the \textsc{MadGraph5\_aMC@NLO} framework
\cite{Degrande:2020evl}, while the fixed-order parton-level program
\textsc{NEWiSH} provides NLO QCD and electroweak (EW) predictions for a wide range of Higgs
decays~\cite{Bellafronte:2026mhp}. Automated one-loop amplitudes in EFTs
are available through \textsc{GoSam} 3.0, which includes example
applications to $h\to b\bar b$ in SMEFT~\cite{Braun:2025afl}.  In addition,
NNLO QCD plus parton shower  predictions for $pp\to Zh\to \ell^+\ell^-b\bar b$ in
SMEFT have been obtained using the MiNNLO$_{\rm PS}$ method, including
the effects of effective Yukawa and chromomagnetic dipole interactions
\cite{Haisch:2022nwz}. 

In this paper we pursue a complementary direction by developing a proof-of-principle implementation of NLO 
SMEFT corrections inside the widely used \textsc{Sherpa}  event-generator 
framework~\cite{Sherpa:2019gpd,Sherpa:2024mfk}. We focus on the benchmark decay $h\to b\bar b$,
 which is a natural testing ground for such an implementation. It is sensitive to the bottom Yukawa interaction, has the largest branching fraction among Higgs decay modes in the SM, and plays a central role in associated-production measurements in which the Higgs boson is reconstructed through $b$-tagged final states~\cite{LHCHXSWG:2016ypw,ATLAS:2024yzu,CMS:2023vzh}.  
 Such measurements motivate event-generator predictions that can be combined with realistic production processes and analysed at the level of differential or fiducial observables.

To define the scope of our work, it is convenient to divide the NLO corrections to $h\to b\bar b$ into three 
gauge-invariant subsets:
\begin{itemize}
\item[(1)] corrections involving virtual or real gluons, which we refer to as QCD corrections;
\item[(2)] corrections involving virtual or real photons, which we refer to as QED corrections;
\item[(3)] virtual corrections involving no photons or gluons, which we refer to as weak corrections.
\end{itemize}
In the SM, the QCD corrections are numerically dominant because they involve the strong rather
than the electroweak coupling.  Assuming no strong hierarchy between the Wilson coefficients 
entering the different types of corrections, the same is true in SMEFT.   Moreover,
from the perspective of event generation, these QCD corrections require adapting the \MCatNLO matching framework~\cite{Frixione:2002ik,Hoeche:2011fd} to Higgs-decay kinematics and matching the result to the \textsc{Sherpa}
QCD parton shower. Their calculation is thus a crucial part of the proof-of-principle implementation 
which carries over directly to other decay processes, and is a main focus of our work.  

Concerning electroweak corrections, the NLO QED terms have an IR structure analogous to their QCD counterparts. We leave their full implementation to future work, where the $h\to b\bar b(\gamma)$ matrix elements could be matched to a QED parton shower along the lines of recent \textsc{Sherpa} developments~\cite{Flower:2026byh,Flower:2026lui}. The NLO weak corrections are more demanding at the amplitude level, owing to the large number of Feynman diagrams and the non-trivial UV-renormalisation required in SMEFT. From the perspective of event generation, however, they are relatively simple, as they are IR finite and contribute only to two-body kinematics. Their inclusion is nevertheless important for global-fit applications, since they probe many Wilson coefficients that should be constrained by data rather than set to zero by hand, and can be especially relevant for processes with no coloured particles at Born level. With a view towards extending the implementation to a broader class of processes, we therefore include the weak corrections in \textsc{Sherpa}, using the matrix elements of~\cite{Gauld:2015lmb,Cullen:2019nnr} and adapting the electroweak input-scheme treatment of~\cite{Biekotter:2023xle}.

The remainder of the paper is organised as follows. In Sec.~\ref{sec:SMEFT} we describe the calculation of
NLO QCD and virtual weak corrections to $h\to b\bar b$ in SMEFT, including the treatment of the bottom-quark mass and electroweak input schemes. In Sec.~\ref{sec:methods} we discuss the matching of the NLO QCD calculation to the parton shower and the procedure used to combine independently generated production and decay
samples. Numerical results for total rates and differential distributions are presented in Sec.~\ref{sec:numerics_ben}, both for Higgs decays in isolation and after embedding them in associated $Zh$ production. We conclude in Sec.~\ref{sec:conclusions}.

\section{$h\to b\bar{b}$ at NLO in SMEFT}
\label{sec:SMEFT}

In this section we cover the calculation of NLO QCD and virtual weak corrections to $h\to b\bar{b}$ in dimension-six SMEFT.  
We write the SMEFT Lagrangian as
$$
\mathcal{L}_{\mathrm{SMEFT}}
=
\mathcal{L}_{\mathrm{SM}}
+\sum_i C_i Q_i,
$$
where $Q_i$ are dimension-six operators in the Warsaw basis \cite{Grzadkowski:2010es}, and the corresponding Wilson coefficients
$C_i$ have mass dimension minus two. Throughout the paper, we retain terms linear in the dimension-six coefficients and neglect contributions quadratic in them. Furthermore, the CKM matrix is approximated as the unit matrix,
and all fermions other than the bottom and top quarks (the latter appearing in weak corrections) are considered massless.   

The QCD corrections involve non-trivial cancellations of IR divergences between real and virtual corrections, as well as subtleties in the definition of the $b$-quark mass, and are thus discussed in some detail. The virtual weak corrections contribute only to the 
two-body decay kinematics. They can therefore be incorporated differentially by multiplying the known
matrix-element corrections for the total decay rate  \cite{Cullen:2019nnr} by the
corresponding two-body phase-space factor; their treatment is summarised at the end of the section.

\begin{figure*}[t]
\centering
\includegraphics[scale=0.5]{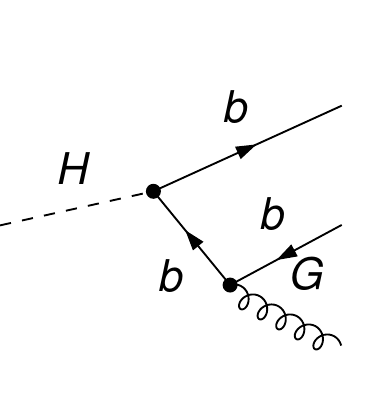}
\includegraphics[scale=0.5]{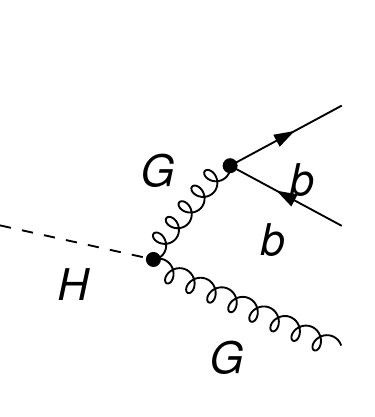}
\includegraphics[scale=0.5]{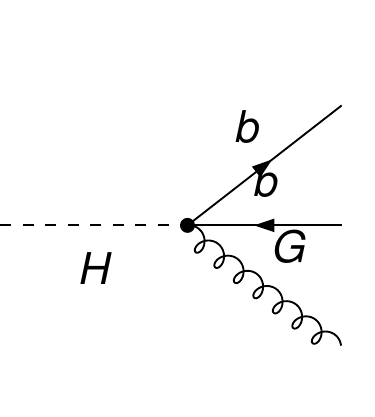}\\
\includegraphics[scale=0.5]{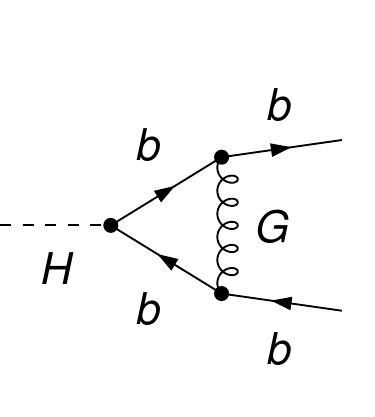}
\includegraphics[scale=0.5]{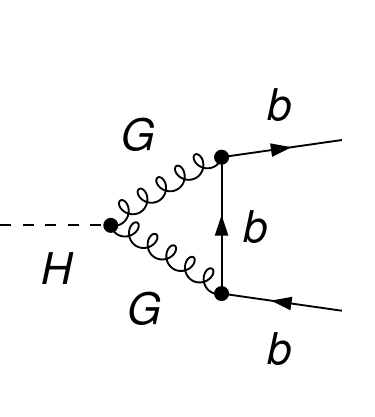}
\includegraphics[scale=0.5]{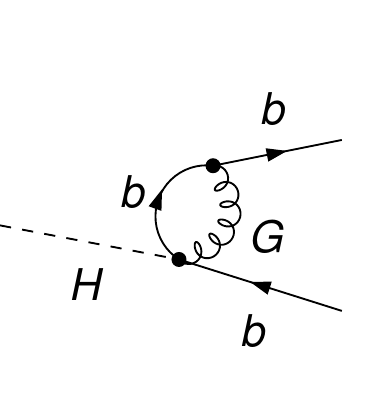}
\caption{Representative real-emission (upper row) and virtual (lower row) contributions to $h\to b\bar b$ at NLO in QCD.
The middle and right diagrams in the upper row are induced by $Q_{HG}$ and $Q_{bG}$, respectively, with the 
corresponding virtual contributions shown below.}
\label{fig:Hbb_FeynDiags}
\end{figure*}

Analytic results for the NLO QCD corrections to the total $h\to b\bar{b}$ decay rate can be found in \cite{Gauld:2016kuu}. The total rate is obtained by combining the UV-renormalised virtual corrections with the real-emission contributions integrated over three-body phase space. The NLO QCD matrix elements receive contributions from the subset of SMEFT operators shown in Table~\ref{tab:QCD_ops}. Typical Feynman diagrams in SMEFT are shown in Fig.~\ref{fig:Hbb_FeynDiags}.   We work in the down-quark mass basis and use the shorthand $C_{bH}\equiv C_{\substack{dH\\ 33}}$ and $C_{bG}\equiv C_{\substack{dG\\ 33}}$, with $Q_{bH}$ and $Q_{bG}$ denoting the corresponding third-generation operators.

From the perspective of event generation, it is useful to separate the SMEFT contributions to differential decay rates into the two broad classes shown in Table~\ref{tab:QCD_ops}.  Operators that modify the SM $hb\bar b$ coupling
inherit the infrared structure of the SM amplitude, so their NLO QCD corrections can be treated with the same 
subtraction and parton-shower matching framework as the SM contribution.  By contrast, the operators
$Q_{bG}$ and $Q_{HG}$ induce vertices not present in the SM.  These are the only operators of this type entering the QCD corrections considered at NLO.

A salient feature of $Q_{bG}$ and $Q_{HG}$ is that the gluon field appears through a field-strength tensor.  
The derivative acting on the gluon field improves the infrared behaviour of the matrix elements, so that
the associated virtual and real-emission contributions are separately infrared finite.\footnote{These terms first enter the decay 
rate at the same perturbative order as the NLO QCD corrections to operators with a non-zero $h b\bar b$ Born amplitude,
and we refer to them as NLO QCD contributions throughout the paper for that reason.
Strictly speaking, however, they are the leading contributions from their
respective Wilson coefficients.} They generate kinematic structures that are not related to the SM contribution
by an overall rescaling and are therefore treated separately in the implementation.  Their contributions to the differential decay rate up to NLO can be obtained by evaluating directly in four space-time dimensions the schematic result
\begin{align}\label{eq:dGamma}
  \done\Gamma
  = \frac{\done\Phi_2}{2 m_H}  \big{|} {\cal M}_{h\to b\bar b}  \big{|}^2
    +\frac{\done\Phi_3}{2 m_H}  \big{|} {\cal M}_{h\to b\bar bg}  \big{|}^2 \,,
\end{align}
where $\done\Phi_j$ is the $j$-body phase-space element, $m_H$ is the  Higgs boson mass,
and it is understood that we retain only  the contribution linear in the corresponding Wilson coefficient.
 
\begin{table}[t!]
\begin{center}
\small
\renewcommand{\arraystretch}{1.5}
\begin{tabular}[t]{c|c}
\hline
$Q_{H\Box}$ & $(H^\dag H)\Box(H^\dag H)$ \\
$Q_{H D}$   & $\ \left(H^\dag D_\mu H\right)^* \left(H^\dag D_\mu H\right)$ \\
$Q_{\substack{dH\\ pr}}$           & $(H^\dag H)(\bar q_p d_r H)$\\
\hline \hline
$Q_{H G}$     & $H^\dag H\, G^A_{\mu\nu} G^{A\mu\nu}$ \\
$Q_{\substack{dG\\ pr}}$        & $ (\bar q_p \sigma^{\mu\nu} T^A d_r) H\, G_{\mu\nu}^A$ \\
\end{tabular}
\end{center}
\caption{\label{tab:QCD_ops}
The dimension-6 SMEFT operators contributing to the NLO QCD corrections to $h\to b\bar{b}$ in SMEFT. The subscripts $p,r$ are flavour indices, and $q_p$ and $d_r$ are left- and right-handed fields, respectively.  The operators in the top part of the table 
modify the $hb\bar{b}$ coupling compared to the SM, while those in the bottom part involve vertices not present in the SM.  
}
\end{table}

The QCD corrections to the bare matrix elements from the remaining operators in Table~\ref{tab:QCD_ops}, 
namely $Q_{bH}$, $Q_{H\Box}$ and $Q_{HD}$, are all proportional to the SM, and in particular share the 
same structure of IR divergences. We deal with these IR divergences using dipole subtraction for massive particles
as developed in \cite{Catani:2002hc}.  In this case, the NLO corrections are obtained by evaluating the schematic formula
\begin{align}
\label{eq:dGamma_dipole}
\Gamma^{(1)} &  =  \int_2 \left[\done\Gamma^{(1)}(h\to b\bar{b})  +\int_1 \done\Gamma_\text{A}^{(1)}(h\to b\bar{b}g)\right]+
 \int_3 \left[\done\Gamma^{(1)}(h\to b\bar{b}g)  - \done\Gamma_\text{A}^{(1)}(h\to b\bar{b}g) \right]    \, ,
\end{align}
where the $(\done)\Gamma^{(1)}$ denotes an NLO correction, and $\done\Gamma^{(1)}_\text{A}$ is a subtraction term that renders the 2- and 3-body contributions individually IR finite.

The (UV-renormalised) squared matrix elements needed to evaluate Eq.~\eqref{eq:dGamma} and Eq.~\eqref{eq:dGamma_dipole}  depend on the renormalisation scheme. As usual, the Wilson coefficients and strong coupling constant $\alpha_s$ are renormalised in the \msbar~scheme, and on-shell wavefunction renormalisation factors are associated with the external legs. The renormalisation of the $b$-quark mass is more subtle.  The phase-space boundaries and IR subtraction terms  are naturally
expressed in terms of the on-shell mass $m_b$.  However, the perturbative series for Higgs decays contains a set of
large logarithms  in the limit $m_b\ll m_H$, which are conventionally absorbed into the running \msbar~mass.  In the present work, we renormalise the $b$-quark mass in the on-shell scheme,  but make use of the parameter
\begin{align}
\label{eq:mb_msbar}
\overline{m}_b \equiv \overline{m}_b(\mu) = m_b\left(1 + \frac{C_F \alpha_s}{\pi}\Delta m_b^{(4,1)} \right) 
\qquad\text{with}\qquad
\Delta m_b^{(4,1)} = -1 -\frac{3}{4}\ln \frac{\mu^2}{m_b^2} \, ,
\end{align}
where $C_F=(N_c^2-1)/(2N_c)$ with $N_c=3$.  Eq.~\eqref{eq:mb_msbar}
corresponds to the NLO relation between the five-flavour $\overline{\rm MS}$
mass $\overline{m}_b$ and the on-shell mass $m_b$.  The on-shell mass is the fundamental input parameter entering the 
renormalisation scheme, phase-space boundaries and subtraction terms, while
$\overline{m}_b$  is used in the overall factor Eq.~\eqref{eq:M2_40_sigma} below as an 
RG-improved normalisation factor that absorbs logarithmic corrections.\footnote{When expressing the 
overall factor in Eq.~\eqref{eq:M2_40_sigma} in terms of $\overline{m}_b$, we expand the pole-to-\msbar~relation consistently through NLO, retaining the corresponding $\mathcal{O}(\alpha_s)$ conversion terms in the NLO coefficients.}

The final ingredient in the renormalisation procedure is the choice of
electroweak input scheme.  This choice affects not only the virtual weak
corrections, but also the Born-level contribution and hence the NLO QCD corrections proportional to it.  
In the current implementation we consider two EW input schemes,
\begin{itemize}
\item the ${\boldsymbol v_\alpha}$ scheme: 
  often called the $\alpha(M_Z)$ scheme and defined by the input set
  $\{\alpha,M_W,M_Z\}$;
\item the ${\boldsymbol v_\mu}$ scheme:
  often called the $G_\mu$ scheme and defined by the input set
  $\{G_F,M_W,M_Z\}$.
\end{itemize}
We streamline the discussion of these schemes by adapting the notation of 
\cite{Biekotter:2023xle}, defining vacuum expectation values in each scheme as
\begin{align}\label{eq:vsigma}
  v_{\alpha} = \frac{2M_W s_w}{\sqrt{4\pi\alpha}}
  \qquad\text{and}\qquad
  v_\mu = \left( \sqrt{2}G_F \right)^{-\frac12}
  \,,
\end{align}
with
\begin{align}
  s_w = \sqrt{1-c_w^2}
  \qquad\text{and}\qquad
  c_w=\frac{M_W}{M_Z}
  \,.
\end{align}
Extending the SM to the SMEFT requires tree-level
shifts from the bare vacuum expectation value of the
Higgs field in the extended theory, $v_T$, to the parameters above.
These are defined through
\begin{align}\label{eq:vT_ren}
  \frac{1}{v_T^2}
  =
    \frac{1}{v_\sigma^2}
    \left[ 1+v_\sigma^2\Delta v_\sigma^{(6,0)} + \dots\right] \,,
\end{align}
where the ellipsis refers to loop corrections to the following tree-level results,
\begin{align}\label{eq:dv_60}
  \Delta v_\alpha^{(6,0)}
  &=\,
    -2\, \frac{c_w}{s_w}
    \left[C_{HWB} + \frac{c_w}{4 s_w}C_{HD} \right]
  \qquad\text{and}\qquad
  \Delta v_\mu^{(6,0)}
  =
    C_{\substack{Hl \\ 11}}^{(3)} + C_{\substack{Hl \\ 22}}^{(3)}  - C_{\substack{ll \\ 1221}} \,.
\end{align}
With this notation, the EW input scheme is fixed by the choice of $\sigma$ in the squared matrix elements.
The choices $\sigma=\alpha$ and $\sigma=\mu$ correspond to the $v_\alpha$ and $v_\mu$ schemes, respectively, with the associated definitions given in Eqs.~\eqref{eq:vsigma} and \eqref{eq:dv_60}.
All Wilson coefficients are defined in the Warsaw basis, following the notation of Table~11 of~\cite{Biekotter:2023xle}.

We end the section by giving explicit results for the squared matrix elements which
form the starting point for event generation.  We write the UV-renormalised 2-body contributions in the form 
\begin{align}
\label{eq:M2_coeff}
 \frac{\big{|} {\cal M}_{h\to b\bar b}  \big{|}^2}{M_{2,\sigma}^{(4,0)}}  & = 1 + v_\sigma^2 K_{2,\sigma}^{(6,0)}  +\frac{1}{v_\sigma^2} W_{2, \sigma}^{(4,1)} + W_{2, \sigma}^{(6,1)} \nonumber \\
&+ \frac{C_F\alpha_s}{\pi}\Bigg[K_2^{(4,1)}\left(1+ v_\sigma^2 K_{2,\sigma}^{(6,0)}\right)
+v_\sigma^2 K_{2,\sigma}^{(6,1)}
 - \frac{1}{\epsilon}\frac{(4\pi)^\epsilon}{\Gamma(1-\epsilon)}\left[1+\frac{1+\beta^2}{2\beta}\ln x\right]\left( 1+v_\sigma^2 K_{2,\sigma}^{(6,0)} \right)\Bigg]  \, ,
\end{align}
where
\begin{align}
  \beta = \sqrt{1-\frac{4m_b^2}{m_H^2}}
  \qquad\text{and}\qquad
  x = \frac{1-\beta}{1+\beta} \,.
\end{align}
Some explanation on the form of the above equation is in order. 
First, the superscript $(i,j)$ denotes a dimension-$i$ contribution at $j$-loop order, 
while the subscript $\sigma$ indicates dependence on the EW  input scheme.\footnote{The notation
makes explicit that the SM  QCD correction $K_2^{(4,1)}$ is scheme independent.}
Second, the quantities $K_2$ denote Born-level and QCD contributions, whereas $W_2$ 
denotes infrared-finite virtual weak corrections.  Finally, the normalisation is chosen relative
to the tree-level matrix element squared in the SM, which we write as
\begin{align}\label{eq:M2_40_sigma}
  M_{2,\sigma}^{(4,0)}
  =
    2 N_c\,\frac{m_H^2}{v_\sigma^2}\,\overline{m}_b^2\,\beta^2
  \,.
\end{align}
The first line of Eq.~\eqref{eq:M2_coeff} contains the Born-level SMEFT correction, which reads
\begin{align}
\label{eq:K20}
K_{2,\sigma}^{(6,0)} = 2 C_{H\Box}-\frac{1}{2}C_{HD} -
 \frac{\sqrt{2} v_\sigma}{\overline{m}_b} C_{bH}  - \Delta v_\sigma^{(6,0)}  \, , 
\end{align}
as well as the one-loop virtual weak corrections, $W^{(i,1)}_{2,\sigma}$,
in both the SM and the dimension-6 SMEFT corrections.  Even in the limit of vanishing $b$-quark mass, 
the explicit results for the weak corrections are rather lengthy. The results in this limit are therefore provided in 
electronic form with the arXiv submission as Mathematica expressions, using \textsc{LoopTools} notation for the one-loop integrals~\cite{Hahn:1998yk}.  The second line of  Eq.~\eqref{eq:M2_coeff} contains the NLO QCD corrections, where
we have made explicit  the IR pole in the dimensional regulator $\epsilon = (4-d)/2$, with $d$ the number of space-time dimensions. This pole cancels against the integrated dipole subtraction terms in Eq.~\eqref{eq:dGamma_dipole}.
Results for these one-loop QCD corrections $K_2^{(i,1)}$, as well as the corresponding real emission terms,
are given in Appendix~\ref{app:NLO_mat}.

\section{Implementation}
\label{sec:methods}

In this section we describe the implementation of the
$h\to b\bar b$ matrix elements derived in the previous section in a
fully exclusive Monte Carlo event-generator framework, which can be used
in realistic analyses aimed at extracting the relevant SMEFT Wilson coefficients.
The squared matrix elements are implemented in a stand-alone library,
\SmeftDecayLib\footnote{%
  \SmeftDecayLib is available from the authors upon request.}.
which we interface with the \Sherpa Monte-Carlo event generator. The NLO QCD
corrections are matched to the  \Sherpa parton shower as described in Sec.~\ref{sec:methods:mcatnlo}.
The resulting fully exclusive Higgs decay events can then be analysed in isolation or combined with an
independently generated simulation of single or multiple Higgs production, as detailed
in Sec.~\ref{sec:methods:production_decay}.
 
\subsection{Parton shower matching}
\label{sec:methods:mcatnlo}

In order to obtain a fully exclusive event description, we match our NLO QCD 
calculation of the differential Higgs decay rate from the previous section to the QCD parton shower
using the \MCatNLO method \cite{Frixione:2002ik}. Within \textsc{Sherpa}~\cite{Sherpa:2019gpd,Sherpa:2024mfk}, 
we match to the \CSShower \cite{Schumann:2007mg} which is based on spin-averaged large-$N_c$ Catani-Seymour dipoles.
The phase-space parametrisation for the integration of the matrix elements is provided by the \Amegic
tree-level matrix element generator \cite{Krauss:2001iv}. Following \cite{Hoeche:2011fd,Hoche:2012tae,Hoeche:2012fm}, the decay width in the \MCatNLO method is defined through
\begin{equation}\label{eq:methods:mcatnlo}
  \begin{split}
    \Gamma^\text{\MCatNLO}
    \,=\;&
	  \int_2\,
	  \overline{\mr{B}}\,\cdot\,
	  \mc{F}_2(t_2)
	  +
	  \int_3
	  \mr{H}\,\cdot\,
	  \mc{F}_3(t_3)\;.
  \end{split}
\end{equation}
Therein, the customarily-denoted $\overline{\mr{B}}$
function of the so-called soft or standard events
is defined as usual as
\begin{equation}\label{eq:methods:bbar}
  \begin{split}
    \overline{\mr{B}}
    \,=\;&
      \done\Gamma^{(0)}(h\to b\bar{b})
      +\done\Gamma^{(1)}(h\to b\bar{b})
      +\int_1
	    \done\Gamma_\mc{F}^{(1)}(h\to b\bar{b}g) \, . 
  \end{split}
\end{equation}
It contains the Born contribution $\done\Gamma^{(0)}$ and
the UV-renormalised one-loop correction $\done\Gamma^{(1)}$ of
Eq.~\eqref{eq:dGamma_dipole}, optionally including
the finite weak corrections, as well as the subtraction terms $\done\Gamma_\mc{F}^{(1)}$.
In contrast to the NLO calculation in the previous section,
however, these subtraction terms are defined through the parton shower $\mc{F}$, detailed below.
The two-parton kinematics of this soft or standard event is
then subjected to the parton shower $\mc{F}_2$,
evolving it from the starting scale $t_2$, which we set
to the Higgs mass in our process.

In general, the parton shower $\mc{F}_n$ evolving an $n$-parton
ensemble from the high scale $t$ using the splitting kernels
$\mr{K}_n$ is defined through
\begin{equation}\label{eq:methods:psmod}
  \begin{split}
    \mc{F}_n(t)
    \,=\;&
      \Delta_n(t_c,t)
      +\int_{t_c}^t\done t'\,
	  \mr{K}_n(t')\,
	  \Delta_n(t',t)\,
      \mc{F}_{n+1}(t')
      \qquad\text{and}\qquad
    \Delta_n(t',t)
    \,=\;
      \exp\left[
        -\int_{t'}^t\done t''\;
         \mr{K}_n(t'')
      \right]
      \,.\hspace*{-20mm}
  \end{split}
\end{equation}
The first term denotes the probability for no emission to
happen between the starting scale $t$ and the infrared
cut-off scale $t_c$ through its associated Sudakov form factor
$\Delta_n(t_c,t)$.
Conversely, the second term describes an emission at scale $t'$
with no other emission between $t$ and $t'$, but is fully inclusive
with respect to any emissions below $t'$.
This region is then further resolved through the iterated parton
shower operating on the newly formed $(n+1)$-parton state
until the infrared cut-off scale $t_c$, of the order of 1\,GeV
marking the transition to non-perturbative QCD dynamics, is reached.

Finally, the hard emission correction is defined as
\begin{equation}\label{eq:methods:h}
  \begin{split}
    \mr{H}
    \,=\;&
      \done\Gamma^{(1)}(h\to b\bar{b}g)
      -\done\Gamma_\mc{F}^{(1)}(h\to b\bar{b}g)
      \;.
  \end{split}
\end{equation}
Its role is to correct the soft-collinear approximation of the
parton shower $\mc{F}_n$ to the full real-emission expression.
As before, this result is inclusive with respect to any emission below the scale
$t_3$ of the one emission already present, being fully resolved
by $\mc{F}_3$.

The SM contribution and the contributions from SMEFT operators that
modify the Born-level $h b\bar b$ amplitude have the same infrared
structure in their real-emission matrix elements.  These are subject to infrared subtraction through
$\done\Gamma^{(1)}_\mc{F}$, and the parton shower $\mc{F}_n$,  determined solely by
infrared physics, remains wholly within the SM.
By contrast, the two- and three-body contributions proportional to $C_{bG}$ and $C_{HG}$ are separately
infrared finite. Their finite two-body terms are included in $\overline{\mr B}$, while their three-body terms are
included directly in $\mr H$, with $\done\Gamma_{\mc F}^{(1)}=0$ for these contributions. 
The finite virtual weak corrections, including those proportional to Wilson
coefficients that first enter at one loop, are likewise included in
$\overline{\mr B}$. All resulting two- and three-parton configurations
are subsequently evolved with the standard QCD parton
shower.

\subsection{Combining production and decay}
\label{sec:methods:production_decay}

To illustrate the use of the Higgs decay implementation in a hadron-collider
environment, we combine independently generated samples for on-shell
Higgs production and Higgs decay. This procedure is appropriate within the narrow-width approximation,
in which production and decay factorise, and the fact that the
Higgs boson is spin-zero ensures the absence of production--decay spin correlations.

In practice, independent production and decay samples are generated
with \textsc{Sherpa} and stored in the HepMC3 format \cite{Verbytskyi:2020sus}. For each
production event, the decay system is first subjected to a random
spatial rotation, distributed uniformly in the Higgs rest frame, and is
subsequently boosted such that its total four-momentum agrees with that
of the Higgs boson in the production event. The on-shell Higgs particle
in the production event record is then replaced by the corresponding
decay system. The random rotation avoids introducing an artificial
correlation between the orientation used to generate the decay event
and the kinematics of the production event. The event weights are then combined such that the 
resulting sample retains the total cross section of the production process multiplied by 
the appropriate differential branching fraction of the decay,
such that the sample is normalised to the appropriate
total cross section,
\begin{equation}\label{eq:methods:production_decay:eventweight}
  \sigma_{\text{prod+decay}}
  = \sigma_\text{prod}^\text{SM}
    \cdot
    \frac{\Gamma_{h\to b\bar{b}}^\text{SMEFT}}
         {\Gamma_{h,\text{tot}}}
    \,.
\end{equation}
The total Higgs boson decay width in the denominator $(\Gamma_{h,\text{tot}})$ is fixed at 4.3~MeV, based on the ATLAS measurement~\cite{ATLAS:2024jry,ParticleDataGroup:2026mpi}, rather than including its SMEFT dependence.
An application of this procedure is presented in Sec.~\ref{sec:Zh_decay_distributions}, in which the
Higgs boson is produced through the $pp\to Zh$ channel.

\section{Numerical results}
\label{sec:numerics_ben}

\begin{table}[t!]
\centering
\begin{tabular}{|cll||cll|}
\hline\hline  
 $M_Z$ & \hc91.188 & GeV & \hl
 $M_W$ & \hc80.379 & GeV  \\
 $m_b$ & \hc\hc4.5 & GeV & \hl
 $m_t$ & 173.21    & GeV  \\
 $m_H$ & 125.0     & GeV & \hl
 $G_F$ & \hc\hc$1.1664 \times 10^{-5}$ & $\text{GeV}^{-2}$  \\
 \hline\hline
 $\alpha(M_Z)$   & $1/128.82$ & & \hl
 $\alpha_s(m_H)$ & \hc\hc$0.11264$ & \\
\hline\hline
\end{tabular}
\caption{%
  Values of the independent input parameters.
  \label{tab:input}
}
\end{table}

In this section we present numerical results for the total decay rate and
differential distributions in $h\to b\bar{b}$ decays implemented
according to Sec.~\ref{sec:methods} in the Monte-Carlo event
generator \Sherpa.
We first discuss the inclusive $h\to b\bar b$ decay rate in Sec.~\ref{sec:tot_rate},
comparing the QCD and weak corrections in the $v_\mu$ and $v_\alpha$ input schemes.
In Sec.~\ref{sec:decay_distributions}, we study Higgs-decay distributions in the Higgs rest frame, focusing on parton-shower effects, the distinct kinematic structures induced by the SMEFT operators, and
the impact of virtual weak corrections. Finally, in Sec.~\ref{sec:Zh_decay_distributions}, we illustrate how the decay
implementation can be combined with associated $Zh$ production at the LHC. 

Throughout the analysis, we use the SM input parameters listed in Tab.\ \ref{tab:input} and set all particle width parameters
entering our matrix elements to zero.
We set the renormalisation scale
to $\mu=m_H$ for all \msbar-renormalised quantities, including the Wilson coefficients, $\alpha_s$, and the running 
$b$-quark mass defined in Eq.~\eqref{eq:mb_msbar}.

\subsection{Total decay rate}
\label{sec:tot_rate}
The total decay rate to NLO in the SM in the $v_\mu$ 
scheme is 
\begin{align}
\label{eq:gam_NLO_vmu}
\Gamma_{v_\mu}^{(4),{\rm NLO}} & = \Gamma^{(4,0)}_{v_\mu}\left(1+ \delta_{\rm QCD}^{(4,1)} + \frac{1}{v_\mu^2} W_{2,\mu}^{(4,1)} \right)
 = \Gamma^{(4,0)}_{v_\mu}\left(1+ 0.207 - 0.009 \right) = 3.02 \, {\rm MeV} \, , 
\end{align}
while in the $v_\alpha$ scheme it is 
\begin{align}
\Gamma_{v_\alpha}^{(4),{\rm NLO}} & = \Gamma^{(4,0)}_{v_\alpha}\left(1+ \delta_{\rm QCD}^{(4,1)} + \frac{1}{v_\alpha^2} W_{2,\alpha}^{(4,1)} \right)
 = \Gamma^{(4,0)}_{v_\alpha}\left(1+ 0.207 - 0.039 \right) = 2.94 \, {\rm MeV} \, .
\end{align}
The QCD corrections are scheme independent while the NLO weak corrections serve to narrow the gap between 
decay rates in the two schemes.  In SMEFT, on the other hand, the dimension-6 contribution in the 
$v_\sigma$ scheme takes the form
\begin{align}
\label{eq:gam_smeft_sigma}
\frac{\Gamma_{v_\sigma}^{(6),{\rm NLO}} }{v_\sigma^2\Gamma_{v_\sigma}^{(4,0)} } & = 
\left[K_{2,\sigma}^{(6,0)} \left(1+\delta_{\rm QCD}^{(4,1)} \right)+ \frac{1}{v_\sigma^2}W_{2,\sigma}^{(6,1)} \right] + 2.55\,C_{HG} + 0.0199\,\frac{m_H^2}{\overline{m}_bv_\sigma}\,C_{bG} .
\end{align}
The last two terms arise from $Q_{HG}$ and $Q_{bG}$,  and as 
discussed in Sec.~\ref{sec:SMEFT} are grouped with the NLO QCD
contributions in our counting, although they are leading for their respective Wilson coefficients.  For
$C_{bG}$ we have kept explicit the scheme-dependent enhancement factor
$m_H^2/(\overline{m}_b v_\sigma)$, which arises because the operator flips chirality
without a mass insertion.  The terms in square brackets contain contributions from Wilson coefficients appearing at Born level, their universal  QCD corrections, and the virtual weak corrections. The weak corrections involve both the Born coefficients and those appearing first at NLO,  of which there are many.  While the QCD corrections to the Born coefficients are universal and the same as in the SM, the weak corrections to them depend on the operator and input scheme. 
Explicitly, in the $v_\mu$  scheme, we have
\begin{align}
\label{eq:weak_vmu}
 K_{2,\mu}^{(6,0)} & \left(1+\delta_{\rm QCD}^{(4,1)} \right)+ \frac{1}{v_\mu^2}W_{2,\mu}^{(6,1)}  = 
  2\left(1 + 0.207 + 0.020 \right)C_{H\Box} -0.5\left(1 + 0.207 + 0.016 \right)C_{HD} \nonumber \\
  &   -1.41\left(1+0.207 + 0 .018 \right)\frac{v_{\mu}}{\overline{m}_b}\,C_{bH} - (1+0.207 + 0.007)\left[C_{\substack{Hl \\ 11}}^{(3)} + C_{\substack{Hl \\ 22}}^{(3)}\right]+\left(1+0.207 - 0.010\right) C_{\substack{ll \\ 1221}} \nonumber \\
&  + \frac{1}{v_\mu^2}\,w_{2,\mu}^{(6,1)} \, ,
\end{align}
where we have written the results for each Born coefficient $C_i$ in the form $c_i\left(1+\delta_{\rm QCD}^{(4,1)} +\delta_{i,\sigma}^{\rm weak}\right)C_i$, so that the three entries in parentheses correspond to the tree-level term, the universal QCD correction, and the operator- and scheme-dependent weak correction, respectively. The quantity $w_{2,\mu}^{(6,1)}$, on the other hand, contains contributions from Wilson coefficients appearing only in the NLO virtual
weak corrections.   The corresponding result in the $v_\alpha$ scheme is
\begin{align}
\label{eq:weak_valpha}
 K_{2,\alpha}^{(6,0)} & \left(1+\delta_{\rm QCD}^{(4,1)} \right)+ \frac{1}{v_\alpha^2}W_{2,\alpha}^{(6,1)}  = 
  2\left(1 + 0.207 + 0.020 \right)C_{H\Box} +1.24 \left(1 + 0.207 - 0.110 \right)C_{HD} \nonumber \\
  &   -1.41\left(1+0.207 + 0.034 \right)\frac{v_{\alpha}}{\overline{m}_b}\,C_{bH}  +3.73 \left(1+0.207 - 0.065\right) C_{HWB} \nonumber \\
&  + \frac{1}{v_\alpha^2}\,w_{2,\alpha}^{(6,1)} \, .
\end{align}
Results for the pure NLO contributions $w_{2,\sigma}^{(6,1)}$  are given in Appendix~\ref{sec:weak_cors}.
We see that in some cases the weak corrections to the  Born-level contributions are sizeable -- for instance, the weak correction to $C_{HD}$ in the $v_\alpha$ scheme is approximately $-11\%$, about half as large as QCD and with the opposite sign.   We will 
study how weak corrections affect shapes of differential distributions after parton showering below.

\subsection{Decay distributions in the Higgs rest frame}
\label{sec:decay_distributions}

\begin{figure}[ht]
    \centering
    % Left column
    \begin{minipage}{0.4\textwidth}
        \centering
        \begin{subfigure}{\linewidth}
            \includegraphics[width=\linewidth]{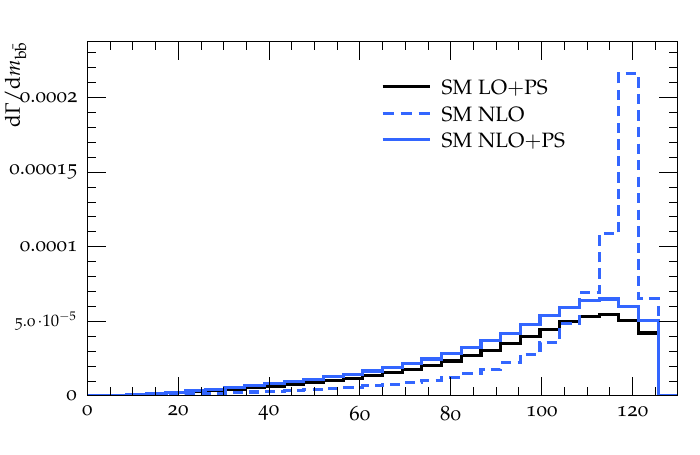} \vskip-0.7cm
             \includegraphics[width=\linewidth]{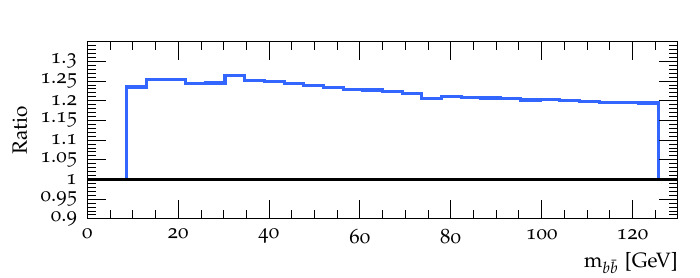}
        \end{subfigure}
        \begin{subfigure}{\linewidth}
            \includegraphics[width=\linewidth]{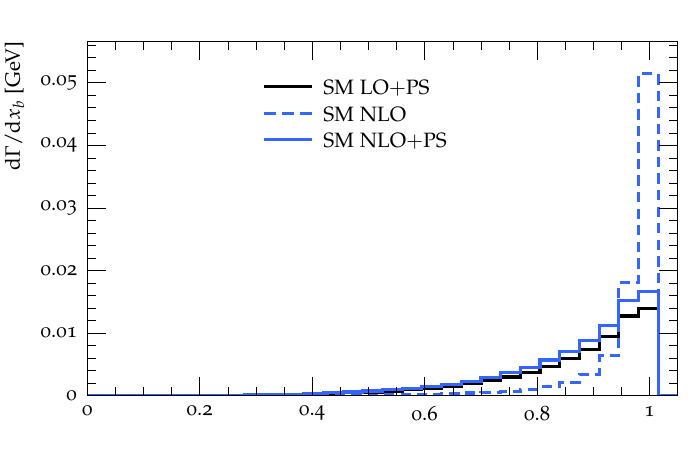}
            \vskip-0.7cm
            \includegraphics[width=\linewidth]{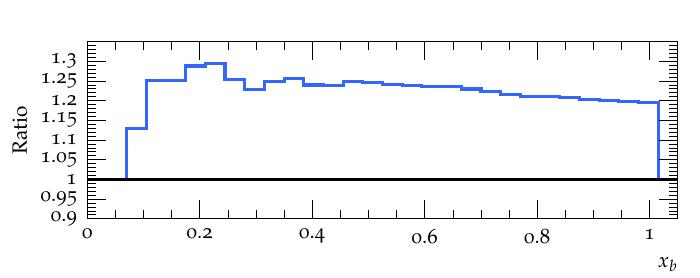}
        \end{subfigure}
    \end{minipage}
    \begin{minipage}{0.4\textwidth}
         \centering
         \begin{subfigure}{\linewidth}
            \includegraphics[width=\linewidth]{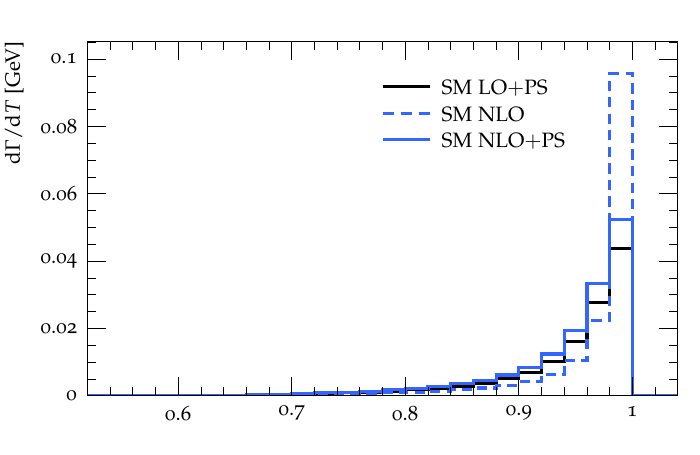}
            \vskip-0.7cm
            \includegraphics[width=\linewidth]{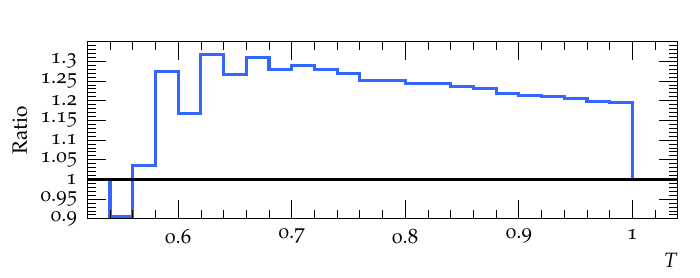}       
             \end{subfigure}
         \begin{subfigure}{\linewidth}
            \includegraphics[width=\linewidth]{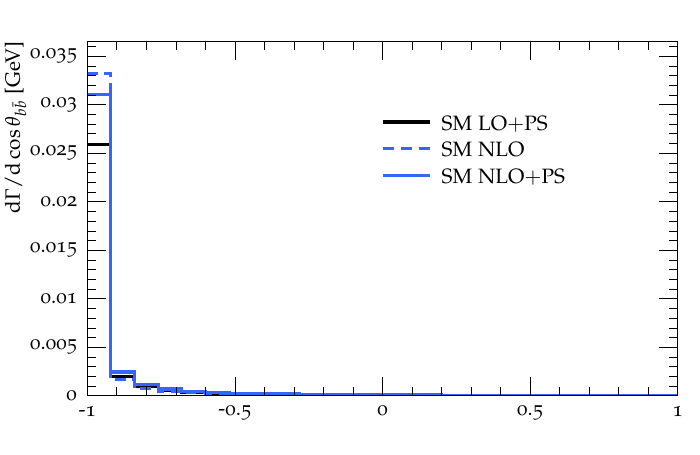}
            \vskip-0.7cm
            \includegraphics[width=\linewidth]{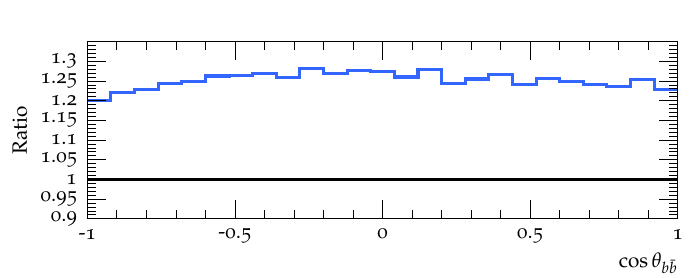}
        \end{subfigure}
          \end{minipage}
          \caption{Differential distributions at various perturbative accuracies within pure QCD, where virtual weak corrections
	and SMEFT Wilson coefficients are set to zero. The lower panels show the ratio of the NLO+PS prediction to LO+PS. 
	\label{fig:sm_ps}}
\end{figure}

\begin{figure*}[t]
	\centering
	\includegraphics[scale=0.45]{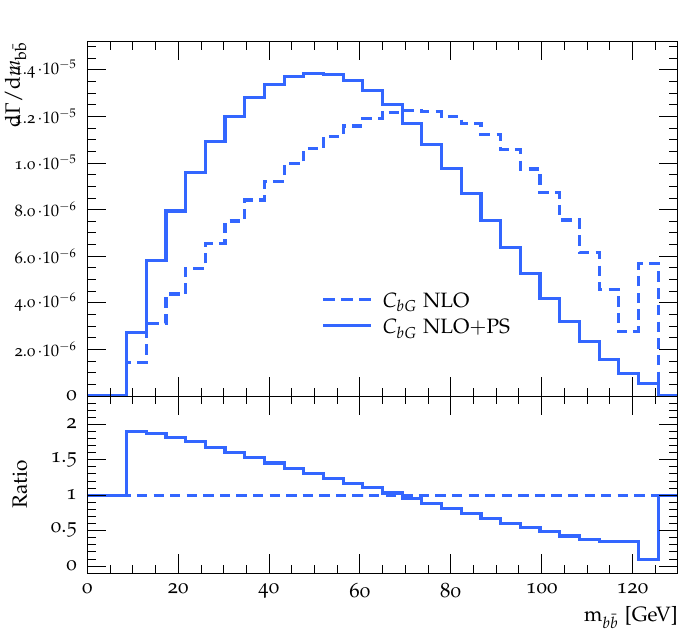}
	\includegraphics[scale=0.45]{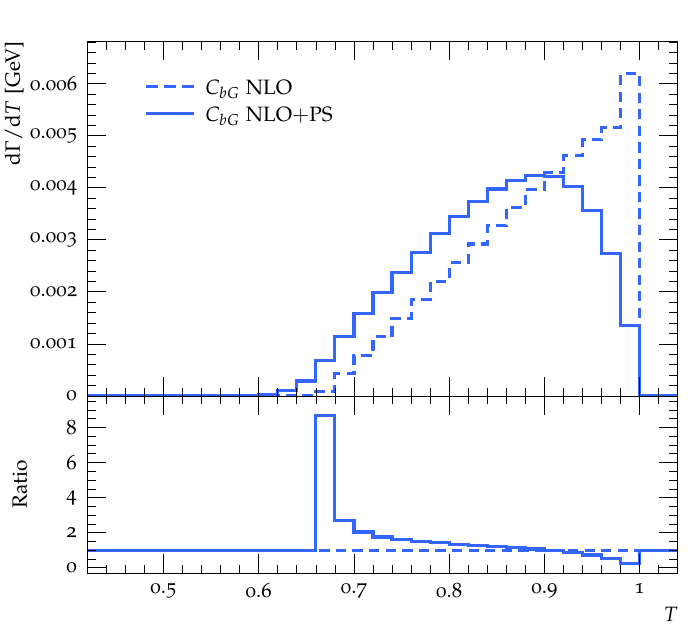}  \\ 
	\includegraphics[scale=0.45]{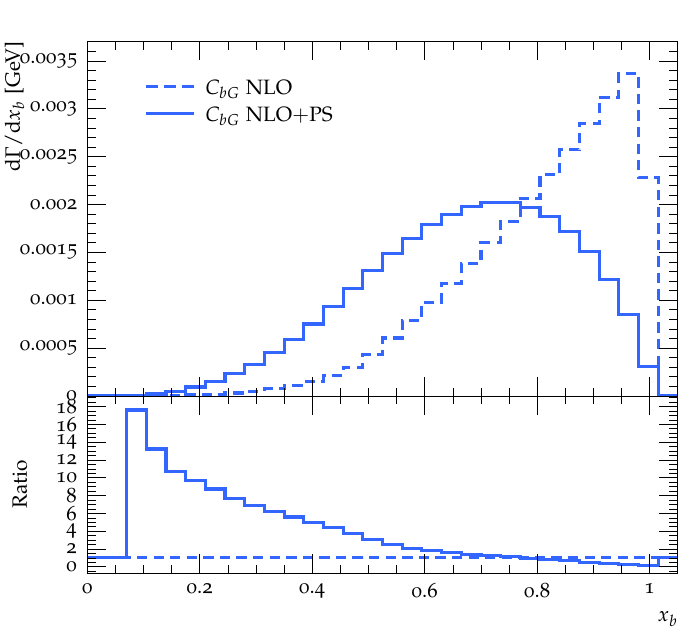}
	\includegraphics[scale=0.45]{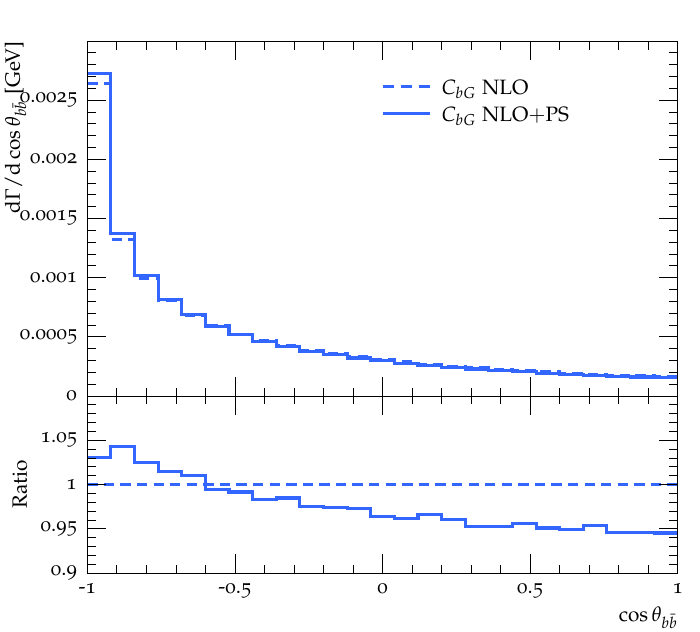}
	\caption{Differential distributions at fixed-order NLO and NLO+PS in QCD for the contribution linear in the Wilson coefficient $C_{bG}$, evaluated with $C_{bG}=1/v_\mu^2$. All other Wilson coefficients and the dimension-four QCD contribution are set to zero. The lower panels show the ratio of the NLO+PS prediction to the fixed-order NLO result. 
	\label{fig:cbg_ps}}
\end{figure*}

\begin{figure*}[t]
	\centering
	\includegraphics[scale=0.45]{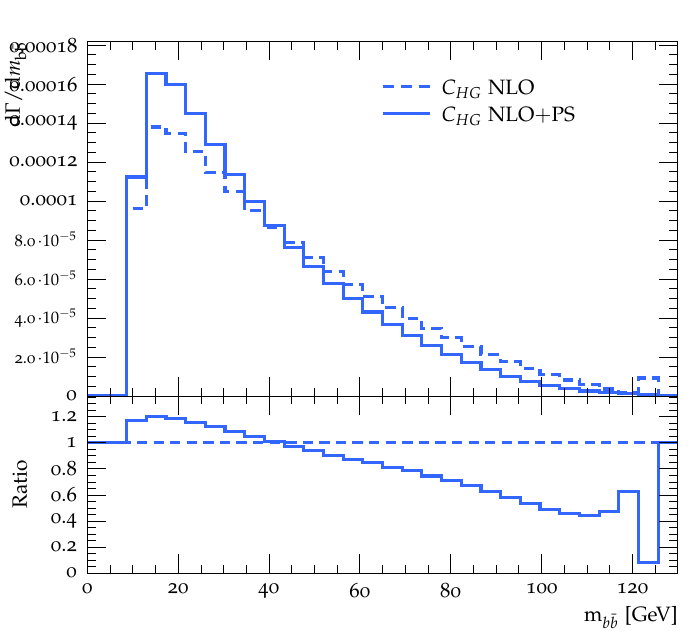}
	\includegraphics[scale=0.45]{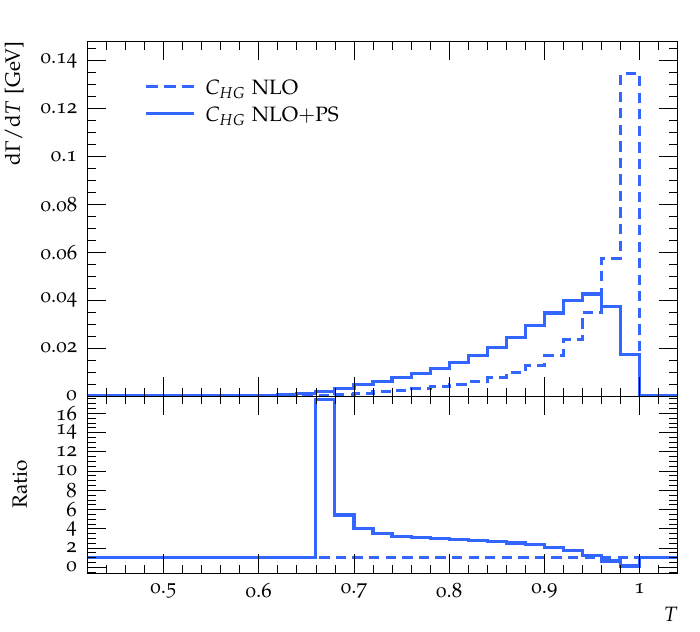}  \\ 
	\includegraphics[scale=0.45]{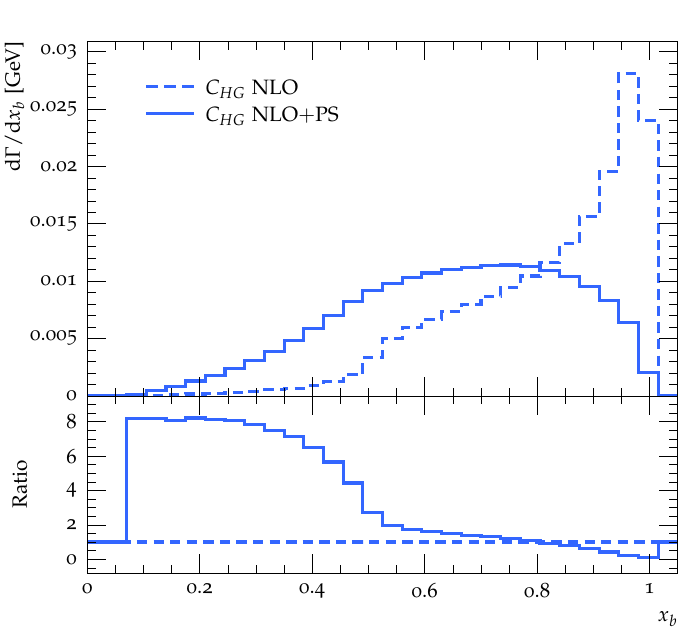}
	\includegraphics[scale=0.45]{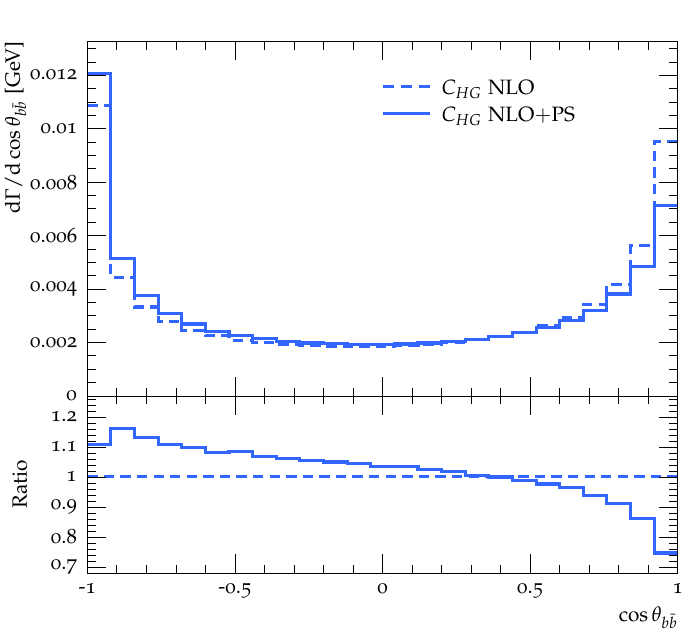}
	\caption{As in Fig.~\ref{fig:cbg_ps}, but for the contribution linear in the Wilson coefficient $C_{HG}$, evaluated with $C_{HG}=1/v_\mu^2$.  
	\label{fig:chg_ps}}
\end{figure*}

\begin{figure*}[t]
	\centering
	\includegraphics[scale=0.45]{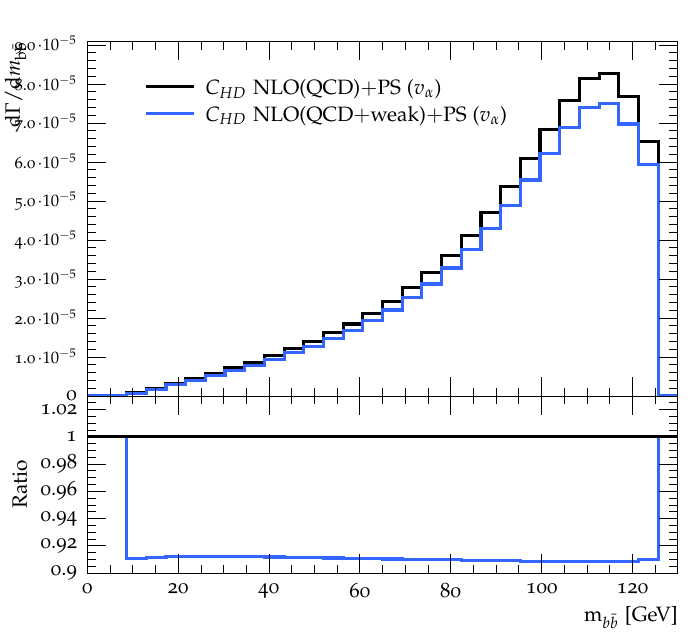}
	\includegraphics[scale=0.45]{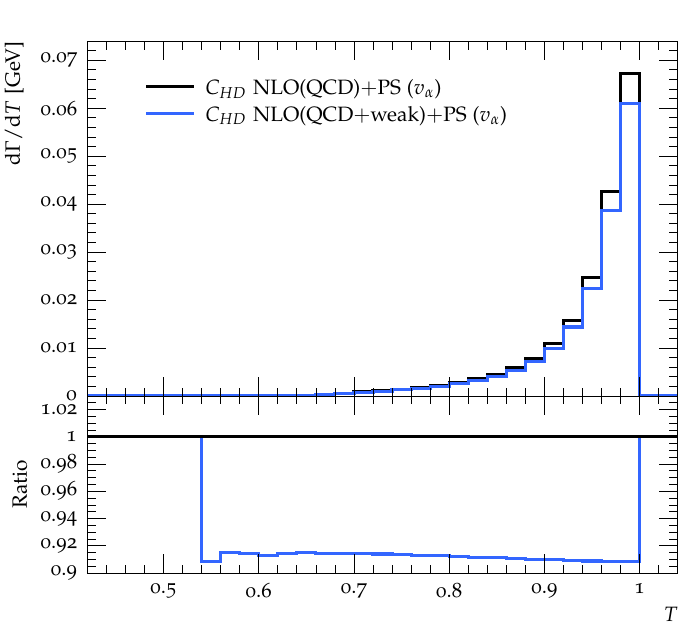}  \\ 
	\includegraphics[scale=0.45]{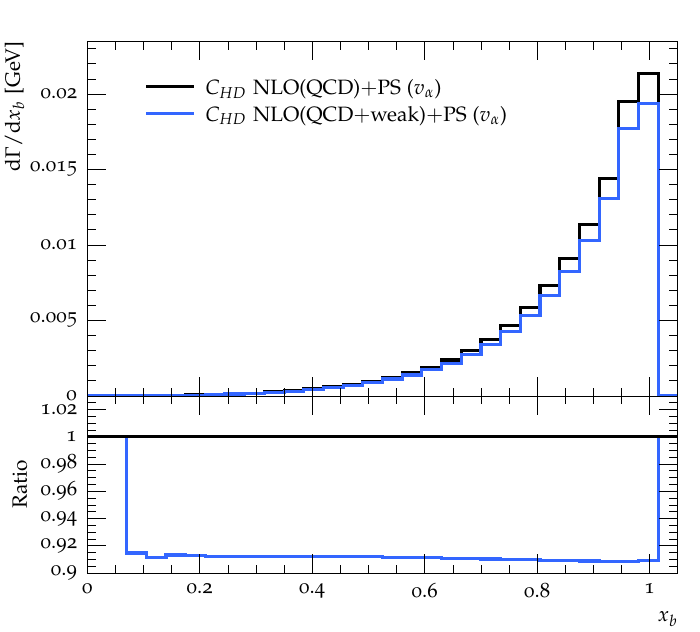}
	\includegraphics[scale=0.45]{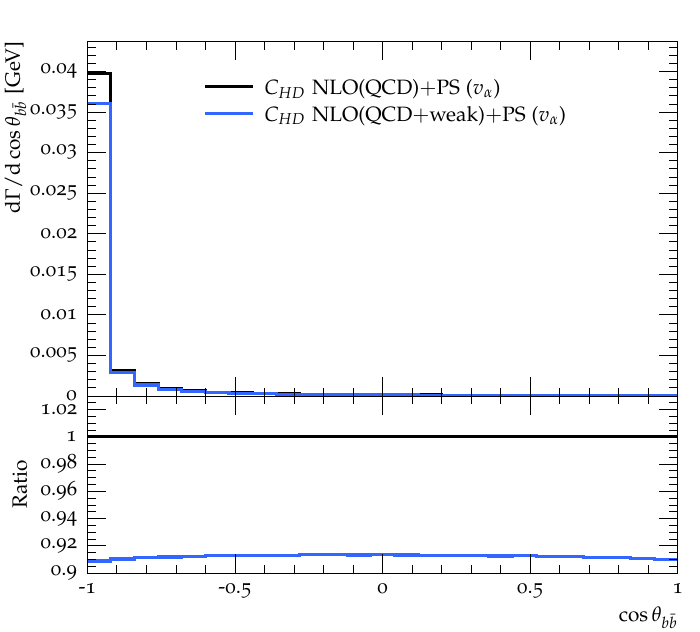}
	\caption{Decay distributions proportional to $C_{HD}$ in the $v_\alpha$ scheme, with $C_{HD}=1/v_\alpha^2$ and all other Wilson coefficients and the dimension-four contribution set to zero. Both predictions are NLO+PS in QCD, with one also including NLO weak corrections. The lower panels show the ratio of the QCD+weak prediction to the QCD-only result.  
	\label{fig:weak_corrections}}
\end{figure*}

In this section we consider decay-level distributions in the Higgs-boson rest frame, working at parton
level before hadronisation.   At NLO in fixed-order there is a unique final-state $b\bar{b}$ pair. 
For the parton-showered samples, the $b\bar b$ system is constructed from the hardest final-state $b$ and  $\bar b$ quarks.  
This prescription avoids imposing an invariant-mass requirement on the $b\bar b$ pair and is suitable for the real-emission contributions, including the $C_{bG}$ and $C_{HG}$ terms. 

We begin with the pure QCD limit of the SM, setting the virtual weak corrections and SMEFT Wilson coefficients to zero. 
We use the $v_\mu$ scheme, which in this case differs from the $v_\alpha$ scheme only through the choice of $v_\sigma$ 
in Eq.~\eqref{eq:M2_40_sigma}. Fig.~\ref{fig:sm_ps} shows differential distributions at LO and NLO matched to the parton shower, labelled LO+PS and NLO+PS, respectively, together with the fixed-order NLO results. The distributions are differential in one of four kinematic variables: the invariant mass of the $b\bar b$ pair, $m_{b\bar b}$; the angle between the $b$ and $\bar b$,
$\cos\theta_{b\bar b}$; the energy fraction of the $b$ quark, $x_b=2E_b/m_H$; or thrust, defined as
\begin{align}
\label{eq:thrust_def}
T = \max_{\hat{\vec n}}
\frac{\displaystyle\sum_i \left|\vec p_i\cdot\hat{\vec n}\right|}
     {\displaystyle\sum_i \left|\vec p_i\right|}\,,
\end{align}
where the sum runs over all final-state partons and $\hat{\vec n}$ is the thrust axis. The fixed-order LO results are not shown, 
as they are delta functions at the Born-level values $m_{b\bar{b}}=m_H$, $\cos\theta_{b\bar b}=-1$, $x_b=1$, or $T =1$
and thus populate only the corresponding endpoint bins. 

The figure shows the expected effects of parton-shower matching, providing a useful consistency check on our implementation. By resumming soft and collinear emissions, which dominate near the Born-level endpoints, the shower softens the fixed-order endpoint enhancements and redistributes events over a broader range of kinematics. This effect is particularly pronounced at LO+PS, where the shower provides the first source of non-trivial radiation.  

A notable feature is that the LO+PS and NLO+PS distributions have
broadly similar shapes, since the evolution kernels of the \CSShower
largely coincide with the respective matrix elements for the decay
of a singlet into a quark-antiquark pair such that the hard emission
corrections added in the NLO matching are small \cite{Schumann:2007mg}.
Their differences are therefore dominated by the overall normalisation associated with the NLO QCD correction to the total rate given in Eq.~\eqref{eq:gam_NLO_vmu}. Some residual shape dependence is visible in all three energy-sensitive observables, while the angular distribution is closest to a pure rescaling.

We now turn to a SMEFT analysis of the same differential distributions, setting aside for the moment operators entering $K_{2,\sigma}^{(6,0)}$ in Eq.~\eqref{eq:K20}, since their NLO QCD contributions are proportional to the corresponding SM contribution and therefore have identical shapes.   In particular, Fig.~\ref{fig:cbg_ps} shows the contribution linear in $C_{bG}$, while Fig.~\ref{fig:chg_ps} shows the corresponding contribution linear in $C_{HG}$, at both NLO and NLO+PS accuracy within the $v_\mu$ scheme. Broadly speaking, the figures reveal two important features.

First, with or without the parton shower, the shapes of the contributions from these two Wilson coefficients differ
substantially both from each other and from the SM.  This behaviour can be understood from the structure of the matrix elements. 
In the case of $C_{bG}$, the virtual corrections in Eq.~\eqref{eq:m2_smeft_qcd} are suppressed by a factor of $m_b^2/m_H^2$ relative to the naive expectation, while in the limit $m_b^2 \ll  t_1, u_1, m_H^2$, the real-emission correction in Eq.~\eqref{eq:m3_smeft_qcd} becomes approximately constant. Consequently, the $C_{bG}$ contributions closely follow the underlying three-body phase-space distribution and are flatter than the SM.  On the other hand, the shapes induced by $C_{HG}$ in Fig.~\ref{fig:chg_ps} are driven mainly by the $g\to b\bar b$ splitting present in the NLO real-emission contribution.
The associated collinear enhancement preferentially produces small opening angles between the $b$ and $\bar b$, 
and therefore enhances the region near the lower kinematic endpoint $m_{b\bar b}\simeq 2m_b$.

Second, as in the SM, the parton shower can also induce sizeable shape-dependent effects for the SMEFT contributions. For 
both $C_{bG}$ and $C_{HG}$ the NLO+PS prediction cannot in general be obtained from the fixed-order NLO result
by a simple overall rescaling.  The effect is most pronounced in observables sensitive to soft or collinear radiation, such as
$m_{b\bar b}$, $T$ and $x_b$, where the ratio panels show substantial redistribution across phase space and resummation effects near kinematic endpoints. Large ratios in certain boundary bins arise when the parton shower populates regions of phase space where the fixed-order prediction has vanishing or strongly suppressed support. A clear example is provided by the thrust distribution: for three-parton kinematics the fixed-order result has a lower boundary at $T=2/3$, while subsequent shower emissions populate the region below this value.  By contrast, the angular distributions are more stable, since soft and collinear radiation primarily modifies energy flow and invariant masses rather than the directions of hard partons.

So far, our study of decay distributions has neglected weak corrections.  At the level of the total rate, the results
in Sec.~\ref{sec:tot_rate} show that these are scheme dependent, introduce many additional Wilson coefficients, and, unlike the QCD corrections, distinguish between the Born-level operators contributing to $K_{2,\sigma}^{(6,0)}$. Consequently, the combined QCD and weak corrections for those operators are no longer simply proportional to the SM contribution.
Weak corrections also modify the shapes of distributions, since in both the SM and SMEFT they contribute to the 
two-body matrix elements but not the three-body ones. To study this effect, Fig.~\ref{fig:weak_corrections} compares the contributions proportional to $C_{HD}$ to the decay distributions in the $v_{\alpha}$ scheme at NLO+PS accuracy in QCD, with and without the inclusion of NLO weak corrections. A notable feature is that the weak corrections change the normalisation by
about 10\% compared to pure QCD, in accordance with the result for the total rate given in Eq.~\eqref{eq:weak_valpha}, 
but have a minimal effect on the shapes.  This is because the weak contribution proportional to $C_{HD}$ has the same shape as the SM LO+PS result, which, as discussed in connection with Fig.~\ref{fig:sm_ps}, closely resembles the corresponding NLO+PS distribution.

\subsection{Higgs decay distributions in associated $Zh$ production at the LHC}
\label{sec:Zh_decay_distributions}

We now illustrate how the decay implementation can be used in a
hadron-collider environment by considering associated $Zh$ production
at the LHC.
In particular, we apply the procedure outlined in Sec.~\ref{sec:methods:production_decay} to study the process
$$
pp\to Zh\to \mu^+\mu^- b\bar b
$$
at a centre-of-mass energy of $\sqrt{s}=13~\mathrm{TeV}$, where the muons arise from $Z$ decay.
To this end, we use \Sherpa \cite{Sherpa:2019gpd} to generate
the process $pp\to Zh$ at NLO QCD accuracy matched to the parton
shower using the \MCatNLO method \cite{Krauss:2001iv,
  Gleisberg:2008fv,Schumann:2007mg,Hoeche:2011fd,Hoche:2012tae,
  Hoeche:2012fm}.
The $Z$ boson is decayed into a muon pair retaining their full
spin-correlations \cite{Hoche:2014kca} and includes higher-order
QED corrections \cite{Schonherr:2008av}.  To simulate the production process, in addition to the parameters
listed above, we use the PDF set \texttt{NNPDF30\_nnlo\_as\_0118}
\cite{NNPDF:2014otw}, interfaced through \LHAPDF \cite{Buckley:2014ana},
and evaluate it at the factorisation scale $\mu_F=m_H$.

Since the aim of this study is to provide a proof-of-principle illustration of the general procedure rather than a precision
phenomenological analysis, we make several simplifying assumptions. First, SMEFT effects are included only in the Higgs decay, while the production process is evaluated in the SM. Second, the analysis is performed at parton level and no jet reconstruction is applied. A realistic particle-level analysis including hadronisation, jet clustering, flavour tagging and QED shower effects in the Higgs decay is left for future work.

All event samples are analysed using \Rivet \cite{Bierlich:2019rhm,Bierlich:2024vqo}.  To define a simple
selection of the leptonic $Z$ decays, we require the two muons to satisfy
$$
p_{T,\mu}>15~\mathrm{GeV},
\qquad
|\eta_\mu|<2.5,
$$
and impose the invariant-mass requirement
$$
75~\mathrm{GeV}<m_{\mu^+\mu^-}<105~\mathrm{GeV}.
$$
The $b\bar{b}$ pair is identified in the Higgs decay event as described above,
and the identities of the two quarks are retained when the decay is embedded
in the production event.  No kinematic cuts are applied to the $b\bar{b}$ system.

\begin{figure*}[t]
	\centering
	\includegraphics[scale=0.45]{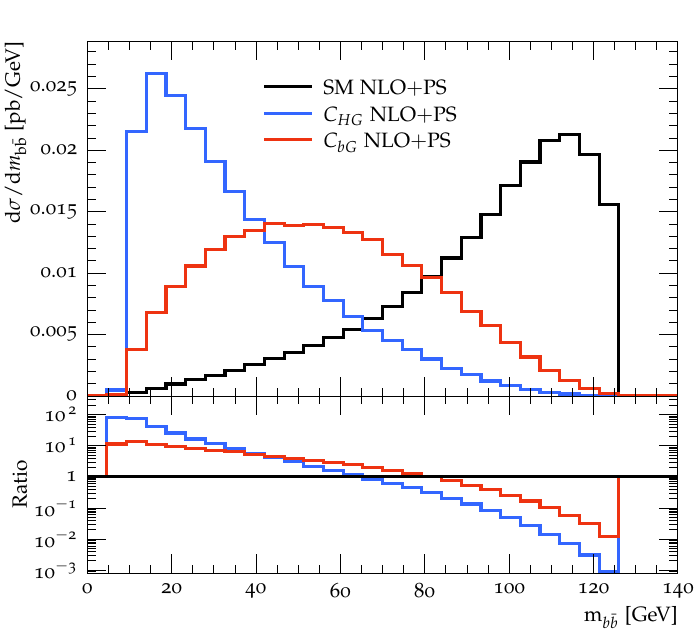} 
	\includegraphics[scale=0.45]{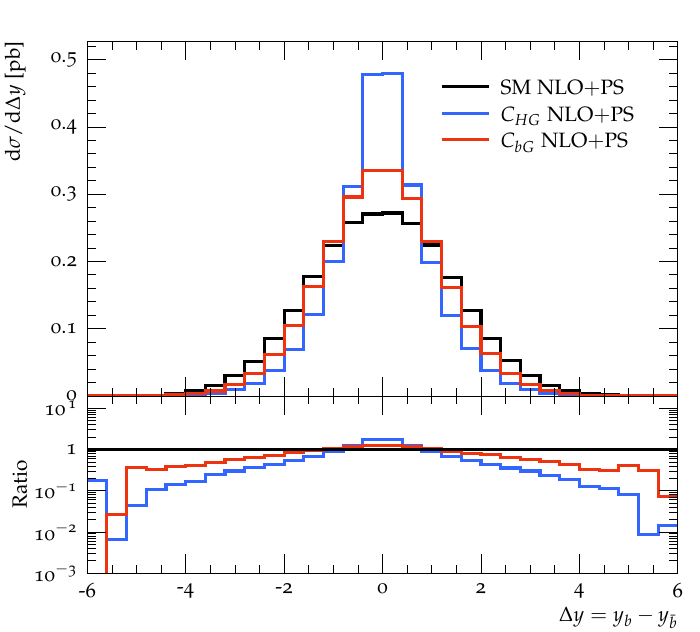} \\
	\includegraphics[scale=0.45]{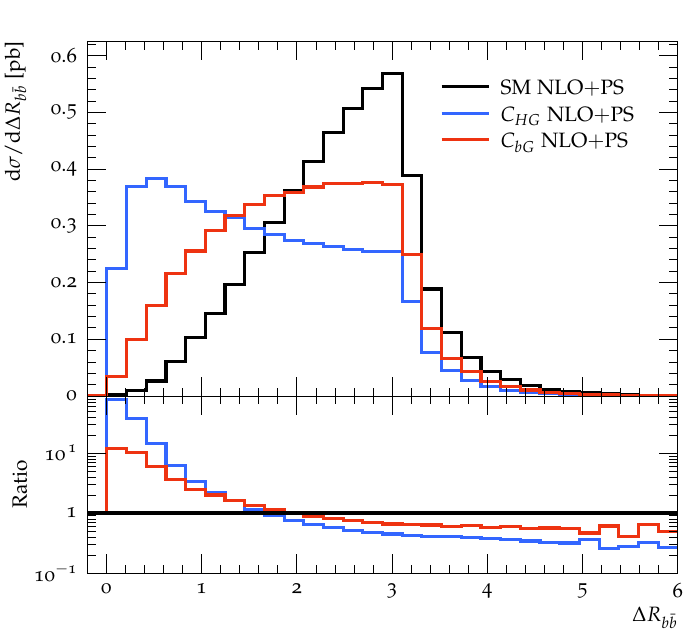}
	\includegraphics[scale=0.45]{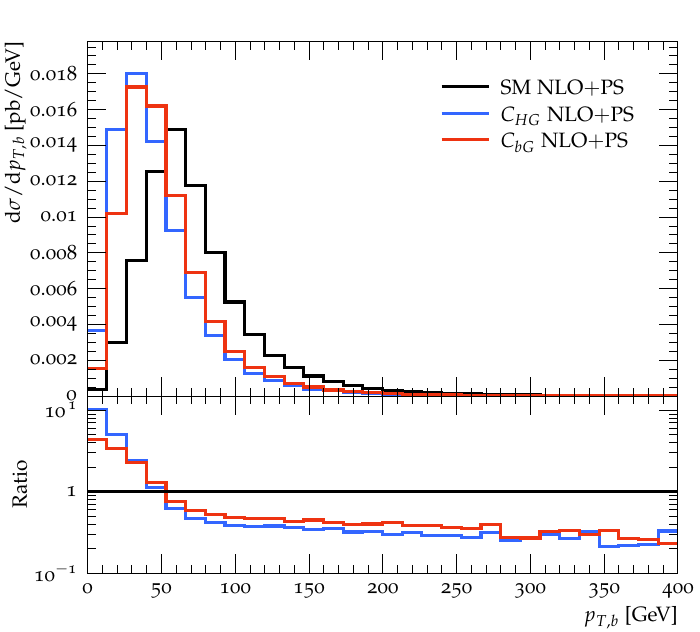} 
	 \caption{The $m_{b\bar b}$, $\Delta y_{b\bar b}$,
$\Delta R_{b\bar b}$ and $p_{T,b}$ distributions in associated $Zh$
production at the LHC with $\sqrt{s}=13~\mathrm{TeV}$. Results are evaluated at NLO+PS accuracy in QCD
using the analysis described in the text. The SM contribution is compared with the contributions linear in
$C_{bG}$ and $C_{HG}$. All results are evaluated in the $v_\mu$ scheme,
with $C_{bG}=C_{HG}=1/v_\mu^2$. The lower panels show the ratios to the
SM prediction. \label{fig:zh_distributions}}
\end{figure*}

Fig.~\ref{fig:zh_distributions} shows the resulting distributions in
$m_{b\bar b}$, $\Delta y_{b\bar b}$, $\Delta R_{b\bar b}$ and $p_{T,b}$,
where $\Delta y_{b\bar b}=y_b-y_{\bar b}$ and
\begin{equation}
\Delta R_{b\bar b}
=
\sqrt{\left(\Delta\eta_{b\bar b}\right)^2+
      \left(\Delta\phi_{b\bar b}\right)^2},
\qquad
\Delta\eta_{b\bar b}=\eta_b-\eta_{\bar b}.
\end{equation}
Here, $y_i$ and $\eta_i$ denote the rapidity and pseudorapidity of particle
$i$, respectively, while $\Delta\phi_{b\bar b}$ is the azimuthal
separation of the two quarks. We set the virtual weak corrections to zero
and consider the SM contribution together with the contributions linear in
$C_{bG}$ and $C_{HG}$, evaluated at NLO+PS accuracy in QCD. Since
$m_{b\bar b}$ is Lorentz invariant, embedding the decay in the production
process leaves its distribution unchanged, confirming that the distinctive
SMEFT-dependent structures found in
Sec.~\ref{sec:decay_distributions} are preserved. The strong preference
of the $C_{HG}$ contribution for small $\Delta y_{b\bar b}$ and
$\Delta R_{b\bar b}$ is the laboratory-frame manifestation of the
collinear $g\to b\bar b$ enhancement discussed in that section. By
contrast, the $C_{bG}$ contribution more closely follows three-body phase
space, and lies between the $C_{HG}$ and SM predictions in these
distributions. The $p_{T,b}$ distribution also distinguishes the three contributions, with
$C_{HG}$ giving the softest spectrum and $C_{bG}$ again intermediate
between $C_{HG}$ and the SM.

\section{Conclusions}
\label{sec:conclusions}

We have presented a \textsc{Sherpa} implementation of NLO corrections in dimension-six SMEFT to the decay $h\to b\bar b$.  
The implementation includes the NLO QCD corrections matched to a QCD parton shower using
an \MCatNLO-type framework for decays.  Virtual weak corrections are included as two-body matrix-element contributions, allowing us to study their dependence on the electroweak input scheme and on Wilson
coefficients that enter first at one loop.  The implementation keeps the treatment of the bottom-quark mass
compatible with massive phase space and infrared subtraction, while
using the running \msbar~mass to absorb certain classes of large perturbative logarithms that appear 
in the small-$m_b$ expansion.

The numerical results illustrate several main features of the implementation.  First, QCD shower effects smear the fixed-order
endpoint structure while preserving the inclusive NLO normalisation. Second, the $C_{bG}$ and $C_{HG}$ contributions produce genuinely different decay shapes, which cannot be reproduced by a rescaling of the SM prediction.  Third, the virtual weak corrections can be important for the normalisation of individual SMEFT directions, especially in the
$v_\alpha$ scheme, but have only a mild impact on the shapes considered here.  Finally, the $Zh$-production 
example at the LHC shows that these decay-level structures can be embedded consistently in a hadron-collider event sample, 
and provides a proof-of-principle demonstration of NLO SMEFT event generation within \textsc{Sherpa}.

There are several natural extensions of this work.  The QED corrections to $h\to b\bar b$ have an infrared structure analogous 
to the QCD corrections and could be incorporated by matching the
$h\to b\bar b(\gamma)$ matrix elements to an appropriate photon-radiation
framework.  A more complete collider study would also include hadronisation, jet clustering, flavour tagging and fiducial selections,
as well as SMEFT effects in the production process.  More broadly, the methods developed here form a basis for extending NLO SMEFT implementations in \textsc{Sherpa} to other decay and production processes,  providing a path towards event-generator predictions for  SMEFT studies with realistic final states and state-of-the-art perturbative accuracy within that framework.

\section*{Acknowledgements}
MS is funded by the Royal Society through a University Research Fellowship
(URF\textbackslash{}R1\textbackslash{}180549, URF\textbackslash{}R\textbackslash{}231031) and Enhancement Awards
(RF\textbackslash{}ERE\textbackslash{}210397,
 RGF\textbackslash{}EA\textbackslash{}181033 and
 CEC19\textbackslash{}100349). MS further acknowledges funding from the STFC Grants
No.\ ST/T001011/1 and ST/P006744/1.

\appendix

\section{Squared matrix elements for QCD corrections}
\label{app:NLO_mat}

In this section we give explicit results for the UV-renormalised two-body and real emission QCD corrections used in
the SMEFT calculation. We begin with the SM contribution. The one-loop QCD correction to the squared matrix element Eq.~\eqref{eq:M2_coeff} reads
 \begin{align}
 K_{2}^{(4,1)} = 
 1+\frac{1}{2}\ln\frac{\mu^2}{m_b^2} - \frac{2\ln x}{\beta}
 + \frac{1+\beta^2}{2\beta}
 \left(2{\rm Li}_2(x) + \frac{1}{2}\ln^2x + 2\ln x+\ln x \ln\left(\frac{\beta^2m_H^2}{\mu^2}\right)+\frac{2\pi^2}{3}     \right) \, .
 \end{align}
One can show explicitly that singular terms in the limit $m_b\to 0$ when $\mu=m_H$ in the above expression are cancelled by the integrated dipole subtraction term Eq.~\eqref{eq:dGamma_dipole}. This cancellation relies on using the \msbar~definition 
$\overline{m}_b$  in the overall $b$-quark mass-squared factor in the tree-level result Eq.~\eqref{eq:M2_40_sigma}, 
and on retaining the NLO terms from the pole-to-\msbar~conversion in Eq.~\eqref{eq:mb_msbar}. This avoids an explicit factor of  $\Delta m_b^{(4,1)}$ from Eq.~\eqref{eq:mb_msbar} and its associated small-mass logarithm in the NLO correction.   

The notation of Eq.~\eqref{eq:M2_coeff} makes clear that the NLO corrections from the SMEFT Wilson 
coefficients appearing in the Born matrix element are proportional to the SM one.  
For the remaining corrections we find 
\begin{align}
  \label{eq:m2_smeft_qcd}
K_{2,\sigma}^{(6,1)} & = C_{HG} 
\left[ 8+6 \ln\frac{\mu^2}{m_b^2} -\frac{1}{\beta}\left(4{\rm Li_2}(-x)+\ln^2x+\frac{\pi^2}{3}\right)  \right]    
\nonumber \\
+ & \frac{C_{bG}}{g_s} \frac{\sqrt{2}m_b}{ v_\sigma} \left[  6 \beta \ln x + 9\ln\frac{\mu^2}{m_b^2}+9  \right] + 2\Delta m_{b,\sigma}^{(6,1)}\, ,
\end{align}
where we have made explicit the NLO SMEFT contribution from $b$-mass renormalisation in 
the on-shell scheme, which is proportional to $C_{bG}$ and is given by
\begin{align}
\Delta m_{b,\sigma}^{(6,1)} = -\frac{C_{bG}}{g_s} \frac{ m_b  }{ \sqrt{2}v_\sigma} \left(  1 + 3 \ln \frac{\mu^2}{m_b^2}\right) \, .
\end{align}
In writing these expressions we have adopted the overall normalisation factor 
$\overline{m}_b^2$ shown in Eq.~\eqref{eq:M2_coeff}. For the SM contribution, and for the SMEFT
operators that contribute already at Born level, this factor is naturally
associated with the bottom Yukawa coupling and serves to absorb logarithms as explained above.
For the interference of the SM amplitude with operators entering first through the QCD correction,
however, only one power of $m_b$ has this origin -- the other arises from a chirality flip. 
The treatment of this second mass factor in the \msbar~scheme is therefore debatable; however,
only a two-loop calculation can resolve it unambiguously, and for simplicity we have chosen 
the above convention. 

The terms in Eq.~\eqref{eq:m2_smeft_qcd} proportional to $C_{bG}$ are suppressed by an 
additional factor of $m_b$ compared to the SM and
$C_{HG}$.   The contribution proportional to $C_{HG}$ itself contains a double logarithm in $m_b/m_H$ in the limit of vanishing $b$-quark mass, as observed already in \cite{Gauld:2016kuu}. For reference, the small-$m_b$ form of Eq.~\eqref{eq:m2_smeft_qcd} is
\begin{align}
K_{2,\sigma}^{(6,1)} &\underset{m_b\to 0}{=} C_{HG}\left[8- \frac{\pi^2}{3}-\ln^2\left(\frac{m_b^2}{m_H^2}\right) +6 \ln\left( \frac{\mu^2}{m_b^2}\right)  \right] \, .
\end{align}
Numerically, there is an accidental cancellation between the large logarithms and other terms, so that in practice the two-body
SMEFT contribution is quite small.

In addition to the one-loop corrections, one must also evaluate the real emission corrections. 
For these three-body corrections, we define SMEFT expansion coefficients as
\begin{align}
\frac{\left| {\cal M}(h\to bbg)\right|^2}{M_2^{(4,0)}} =  \frac{C_F\alpha_s\pi}{m_H^2}   \left( 
K_{3}^{(4,1)} \left[1+ v_\sigma^2 K_{2,\sigma}^{(6,0)} \right]
+ v_\sigma^2 K_{3,\sigma}^{(6,1)}\right) \,,
\end{align}
In quoting results, it is convenient to define the Mandelstam
variables
\begin{align}
s = (p_b+p_{\bar{b}})^2 \, , \quad t_1 = (p_b + p_g)^2-m_b^2 \,,  \quad u_1 = (p_{\bar{b}} + p_g)^2-m_b^2 \,  ,
\end{align}
which satisfy the relation
\begin{align}
 s+t_1+u_1=m_H^2 \,.
\end{align}
In terms of these variables, the three-body contribution in the SM reads
\begin{align}
  \label{eq:m3_sm_qcd}
K_{3}^{(4,1)}  =\frac{ 8}{\beta^2}
&\bigg[
 \frac{8m_b^4 - 2m_b^2\left(s+u_1 -t_1\right) + t_1 u_1   }{t_1^2}
+  \frac{8m_b^4 - 2m_b^2\left(s+t_1 -u_1\right) + t_1 u_1   }{u_1^2} \nonumber \\
&+ \frac{16m_b^4  - 4m_b^2\left(3s+2t_1+2u_1 \right) + 2 \left(s+t_1\right)\left(s+u_1\right)}{u_1t_1}
\bigg] \, .
\end{align}
For the SMEFT result not proportional to the SM we find
 \begin{align}
   \label{eq:m3_smeft_qcd}
K_{3,\sigma}^{(6,1)} =  \frac{32}{\beta^2} &\bigg\{ 
C_{HG}  \left[\frac{-4m_b^2\left(t_1+u_1\right) +2 t_1 u_1 + 2 s t_1 + t_1^2+u_1^2 }{s t_1} 
+\left\{t_1\leftrightarrow u_1 \right\} \right] \nonumber \\
& + \frac{C_{bG}}{g_s}\frac{m_H^2}{\sqrt{2}\overline{m}_b v_\sigma}
\left[1 - \frac{m_b^2}{m_H^2}\frac{\left(t_1+u_1\right)^2}{t_1 u_1}  \right]
\bigg\} \,.
\end{align}

\section{Weak corrections from operators appearing first at NLO}
\label{sec:weak_cors}
Here we give results for the pure NLO weak corrections, using the numerical inputs in Tab.~\ref{tab:input} and $\mu=m_H$. In the $v_\mu$ scheme they are
\begin{align}
\frac{1}{v_\mu^2}w_{2,\mu}^{(6,1)} & =
\bigg(
-5.7 \frac{v_\mu}{\overline{m}_b} C_{\substack{dW \\ 33}}
-4.6 C_{\substack{uW \\ 33}}
-3.1 C_{\substack{Hq \\ 33}}^{(3)}
-3.0 C_{\substack{uH \\ 33}}
+3.0 \frac{v_\mu}{\overline{m}_b} C_{\substack{quqd \\ 3333}}^{(1)}
\nonumber \\ &
+2.4 C_H
-2.2 C_{\substack{ll \\ 1122}}
+2.0 C_{HW}
-1.3 C_{\substack{qd \\ 3333}}^{(8)}
\bigg)\times10^{-2}
\nonumber \\
&+ \bigg(
-9.8 C_{\substack{qd \\ 3333}}^{(1)}
+5.9 C_{\substack{Hq \\ 33}}^{(1)}
+5.7 \frac{v_\mu}{\overline{m}_b} C_{\substack{quqd \\ 3333}}^{(8)}
+4.1 C_{HWB}
-3.5 \frac{v_\mu}{\overline{m}_b} C_{\substack{Hud \\ 33}}
\nonumber \\ &
+2.9\left[
C_{\substack{lq \\ 1133}}^{(3)}
+C_{\substack{lq \\ 2233}}^{(3)}
\right]
-2.2 C_{\substack{Hd \\ 33}}-1.0\left[ C_{\substack{Hl \\ 11}}^{(1)}
+C_{\substack{Hl \\ 22}}^{(1)} \right]
\bigg)\times10^{-3}
\nonumber \\ &
+ \bigg(
9.0 C_{HB}
+2.8 \frac{v_\mu}{\overline{m}_b} C_{\substack{dB \\ 33}}
\bigg)\times10^{-4}\, .
\end{align}
In the $v_\alpha$ scheme
\begin{align}
\frac{1}{v_\alpha^2}w_{2,\alpha}^{(6,1)} & =
-0.12 C_{\substack{Hq \\ 33}}^{(3)}
+\bigg(
-7.9 C_{\substack{Hu \\ 33}}
-6.0 \frac{v_\alpha}{\overline{m}_b} C_{\substack{dW \\ 33}}
+5.8 C_{\substack{Hq \\ 33}}^{(1)}
-4.0 C_{\substack{uW \\ 33}}
\nonumber \\ &
+3.1 \frac{v_\alpha}{\overline{m}_b}
C_{\substack{quqd \\ 3333}}^{(1)}
-3.1 C_{\substack{uH \\ 33}}
-3.0 C_{\substack{uB \\ 33}}
+2.4 C_H
+1.9 C_{HW}
-1.3 C_{\substack{qd \\ 3333}}^{(8)}
-1.0 C_{\substack{qd \\ 3333}}^{(1)}
\bigg)\times10^{-2}
\nonumber \\
&+\bigg(
-9.3\bigg[
C_{\substack{Hq \\ 11}}^{(3)}
+C_{\substack{Hq \\ 22}}^{(3)}
\bigg]
-8.5\bigg[
C_{\substack{Hu \\ 11}}
+C_{\substack{Hu \\ 22}}
\bigg]
-6.4 C_W
+6.0 \frac{v_\alpha}{\overline{m}_b}
C_{\substack{quqd \\ 3333}}^{(8)}
\nonumber \\ &
+4.2\bigg[
C_{\substack{Hd \\ 11}}
+C_{\substack{Hd \\ 22}}
+C_{\substack{He \\ 11}}
+C_{\substack{He \\ 22}}
+C_{\substack{He \\ 33}}
+C_{\substack{Hl \\ 11}}^{(1)}
+C_{\substack{Hl \\ 22}}^{(1)}
+C_{\substack{Hl \\ 33}}^{(1)}
-C_{\substack{Hq \\ 11}}^{(1)}
-C_{\substack{Hq \\ 22}}^{(1)}
\bigg]
\nonumber \\ &
-3.6 \frac{v_\alpha}{\overline{m}_b}
C_{\substack{Hud \\ 33}}
-3.1\bigg[
C_{\substack{Hl \\ 11}}^{(3)}
+C_{\substack{Hl \\ 22}}^{(3)}
+C_{\substack{Hl \\ 33}}^{(3)}
\bigg]
+2.0 C_{\substack{Hd \\ 33}}
\bigg)\times10^{-3}
\nonumber \\ &
+\bigg(
-7.2 C_{HB}
+2.9 \frac{v_\alpha}{\overline{m}_b}
C_{\substack{dB \\ 33}}
\bigg)\times 10^{-4}\, .
\end{align}
In both cases we have kept enhancement factors of $v_\sigma/\overline{m}_b$ symbolic.

%= bibliography ===================================
\bibliographystyle{amsunsrt_mod}
\bibliography{references}

%= end ============================================
  \end{document}